# Smartphone as a Personal, Pervasive Health Informatics Services Platform: Literature Review

K. Wac
University of Geneva, Switzerland, Institute of Services Science, Quality of Life Group, Geneva, Switzerland

### Summary

*Objectives*: The article provides an overview of current trends in personal sensor, signal and imaging informatics, that are based on emerging mobile computing and communications technologies enclosed in a smartphone and enabling the provision of personal, pervasive health informatics services.

*Methods*: The article reviews examples of these trends from the PubMed and Google scholar literature search engines, which, by no means claim to be complete, as the field is evolving and some recent advances may not be documented yet.

*Results*: There exist critical technological advances in the surveyed smartphone technologies, employed in provision and improvement of diagnosis, acute and chronic treatment and rehabilitation health services, as well as in education and training of healthcare practitioners. However, the most emerging trend relates to a routine application of these technologies in a prevention/wellness sector, helping its users in self-care to *stay* healthy.

*Conclusions*: Smartphone-based personal health informatics services exist, but still have a long way to go to become an everyday, personalized healthcare-provisioning tool in the medical field and in a clinical practice. Key main challenge for their widespread adoption involve lack of user acceptance striving from variable credibility and reliability of applications and solutions as they a) lack evidence-based approach; b) have low levels of medical professional involvement in their design and content; c) are provided in an unreliable way, influencing negatively its usability; and, in some cases, d) being industry-driven, hence exposing bias in information provided, for example towards particular types of treatment or intervention procedures.

### Keywords

Personal health services, wireless technology, information services, sensors, automatic data processing, personal health informatics applications, physiology, telemedicine, personalized medicine, mobile phone, body area network

## 1 Introduction

A ubiquitous availability of personal mobile devices and high-capacity wireless networks enable innovative applications in different aspects of our daily life, *e.g.*, communication, education or entertainment. Over the past decade, particularly smartphones become more prevalent, *e.g.*, 50% of U.S. mobile phone users had a smartphone at the end of 2011 [1] and 300 million smartphones were being sold worldwide in 2010 [2] and 500 million are predicted for 2012 [3]. At the end of the 2011, there were more smartphones sold in the US, than PCs [3]. A smartphone is different from (low-end) feature phone, as the former has larger computing power and storage capabilities, as well as a set of advanced features (camera, touch-screen) and advanced application programming interfaces (APIs) for running third-party data-based applications (*apps*), while the latter is mainly designed for voice and text-based interactions and come with a pre-installed set of fixed applications and limited features.

The smartphone market leaders include Apple's iPhone, RIM's Blackberry, Google's Android and Microsoft Windows Mobile platforms, each having own unique set of features [4] and applications distribution channel, such as app-store (Apple) or an app-market (Google). There has been a large increase in the number of apps downloaded on smartphones over the past years, with 300 million applications downloaded in 2009, 5 billion in 2010 and 12 billion in 2011 [5]. Besides the dedicated smartphone apps, users can access web-based content on their smartphone, *i.e.*, websites with information, gaming applications or positioning and navigation services. Furthermore, as we have investigated in one of our large population studies, people carry their smartphones around almost all the time with them [6].

In parallel, an increasing availability of smartphone built-in (*e.g.*, magnetometer, accelerometer, air pressure), as well as external sensors, capturing different phenomena (*e.g.*, electric, magnetic, electrochemical, mechanical, thermal and optical), enable a development of new sensor systems for measurements of a state of a phone user and his/her surrounding environment. Particularly there exist sensor systems worn on the body or around the body forming a *Body Area Network* (BAN), for which a smart-phone is a central processing unit. These BANs can measure user's psychophysiology and environmental conditions, integrating for example heart rate, respiration rate, and body and, as well as ambient temperature sensors. These BANs and associated mobile applications can be used as a tool for gathering quality data for medical research, or regular healthcare practice, as data can be gathered from the subjects unobtrusively for long periods of time, in a laboratory, as well as in a subject's natural environments.

This article reviews some examples of trends in personal sensor, signal and imaging informatics from the PubMed and Google scholar literature search engines, without claiming completeness, as the field is continuously enriched with new developments. We recall that, according to definitions by Lehmann, Aach and Witte [7], *sensor informatics* refers to an acquisition process of data from physical sensors, *signal informatics* – to data management, visualization, and simple data analysis, while *imaging informatics* refers to processing data in two or more dimensions. *Personal health informatics* refers to usage of such processed data in the health(care) processes that are designed to meet the particular needs and situation of the care receiver [7]. We expand this definition twofold. Firstly, we expand it by including care provider, *e.g.*, clinician, using such processed data for own training or in the health(care) processes that are designed to making the decisions upon care receiver's diagnosis, interventions, treatments, or rehabilitation. Moreover, we expand this definition towards receiver's self-care and wellness and prevention processes.

This article is structured as follows. Section 2 presents details of the method employed in our research, while Section 3 presents its results. Section 4 summarizes the findings and discusses their limitations, while Section 5 concludes this article.

## 2 Methods

For the purpose of this overview, we scope our research as follows. We consider smartphones as portable mobile devices that have computing power, storage and interactive wireless data communications capabilities, and are able to run data-based software applications. Applications provided exclusively on desktop personal computers, notebook (laptop) computers, pagers or handheld calculators are not considered in our review. Moreover, considered application must aim to improve or promote health or health service use and quality. These include applications designed to improve primary or secondary prevention, diagnosis, acute or chronic treatment, rehabilitation or practitioners' training. We therefore include applications provided on smartphones owned or directly used by a patient, practitioner or a lay-person. Furthermore, we include studies that have a controlled design to evaluate the application as a primary component under evaluation, including both randomized controlled trials and non-random group allocations, embracing also numerous feasibility studies. We do not impose a limit on study participants in terms of age, gender, ethnicity and morbidities (for patients) and staff role and occupation (for practitioners, *e.g.*, surgeon, psychologist). Also, we are including all outcome measures, both objective (*e.g.*, performance) and self-reported, with the emphasis on user-acceptability factors (*e.g.*, usability, usefulness).

The results presented in this article are based on the relevant literature search using PubMed [8] search engine for scientific, peer-reviewed literature, enclosing majority of relevant for us articles from the Association for Computing Machinery (ACM) [9] and Institute of Electrical and Electronics Engineers (IEEE) [10] digital libraries and their search engines. Additional results available in the latter search engines are omitted from our study as these relate to results of simulations and/or small feasibility prototypes, in most cases without an involvement of real (health)care practitioners or patients. The search terms used are „smartphone", „cell phone" or „mobile phone" technologies (*i.e.*, the keyword must have appeared in the paper title or in an abstract) and the publication year was 2010-2012. We would like to notice that the articles that relate to the effects of electromagnetic radiation, as well as articles evaluating smartphones use influence on safety while multitasking, *i.e.*, driving, walking or conducting other daily life activities, were not considered. For the purpose of presenting the recent advances and developments documented in prevention/wellness sector, we have run a query in the Google search engine [11]; as majority of these advances are not yet documented scientifically.

The search was not exhaustive, but rather informative, where the main goal was to get an overview of current trends in personal sensor, signal and imaging informatics and map them on current developments in mobile computing and communications technologies enclosed in a smartphone. The selection of the references cited in this article is based on their novelty and possible contribution into the field of the personal pervasive health informatics services. In total, 772 articles were found and around 100 are selected for this article, given the inclusion criteria described above and the fact that for some objectives, *e.g.*, telemonitoring of vital signs there were many examples of similar studies, and to avoid repetition, only the most recent studies have been selected to be included in this overview.

## 3 Towards Smartphone-Based Personal, Pervasive Health Informatics Services

This section presents the results of the literature search. First, we present advances in the surveyed technologies, as employed in provision of diagnosis (Section 3.1), treatment (Section 3.2) and rehabilitation (Section 3.3) healthcare services, and evaluated for their feasibility, performance and usability and usefulness with healthcare practitioners and their patients. Second, we present examples of employing smartphone in training of practitioners in different domains (Section 3.4). Finally, we present advances and developments documented in prevention/wellness sector (Section 3.5).

### 3.1 Diagnosis

There are multiple examples of employing smartphone applications towards improving diagnosis process. For instance, Qiao, *et al.* [12] develop a smartphone-based measurement method for reliable and efficient quantitative assessment of the curvature in diagnosis of Scoliosis, by facilitating the measurement of the Cobb angle, *i.e.*, the angle between two lines, drawn perpendicular to the upper endplate of the uppermost vertebrae involved and the lower endplate of the lowest vertebrae involved. Rigoberto *et al.* [13] successfully prove a possibility of using a smartphone as a tool for measuring anticipatory postural adjustments in healthy subjects (before the beginning of normal gait in healthy subjects). Yamada *et al.* [14, 15] develop a smartphone application that uses build-in accelerometer to objectively assess abnormal gait in patients with rheumatoid arthritis (RA). In the consecutive studies, they same authors assess the use of smartphone while walking, indicating that the measure of dual-tasking ability, and particularly changes in gait, are effective measure for a risk of fall assessment. Similarly, Lee and Carlisle [16] focus on recognition of falls (forwards, backwards, lateral left and lateral right) by elderly, using the built-in smartphone accelerometer. Moreover, Duchene and Hewson [17] positively evaluate usability and acceptability of a modified bathroom scale, which, connected to smartphone, attempted to quantify balance of older adults. Shin *et al.* [18] use smartphone to reliably measure shoulder range of motion; the authors emphasize convenience and cost-effectiveness of this method comparing to the gold standard methods.

Engel *et al.* [19] enable a remote, real-time monitoring of free flaps via smartphone photography over the third generation of mobile telephony (3G) wireless network. A prospective study evidences, that diagnostic accuracy is comparable with the accuracy rate for in-person examinations for free flap monitoring, and the response time is shorter. Similarly, Sprigle *et al.* [20] develop an accurate and reliable wound assessment technique, using camera of a smartphone. Lamel *et al.*, [21] evaluate mobile tele-dermatology and claim that cellular phones present an innovative and convenient modality of providing dermatologic consultations for skin cancer screening.

Choi *et al.* [22] assess an accuracy of interpretation of remotely downloaded pocketsize cardiac ultrasound images on a web-enabled smartphone, validated against workstation evaluation. The results show that remote expert echocardio-graphic interpretation can provide backup support to point-of-care diagnosis by non-experts when read on a smartphone. Furthermore, mobile-to-mobile consultation may improve access in previously inaccessible locations to accurate echocardiographic interpretation by experienced cardiologists. Similarly, Crawford *et al.* [23] positively assess that the telementored „just-in-time" telesonography of the Focused Abdominal Sonography for Trauma (FAST) protocol over a smartphone and indicate that the picture quality enables diagnostic quality by the remote experts.

Parakh and Chaturvedi [24] present innovative way of documenting tachycardia in rural India; the patient records video of neck pulsation done with the help of mobile phone camera, which can be then evaluated reliably by a trained physician. Similarly, Althubaiti *et al.* [25], use pictures gathered the phone for effective diagnosis and management of acute and sub-acute problems in hand surgery. The authors emphasize that cell phone video technology is available, cheap, portable, and can be used to help improving consultation system in general and reduce the costs of unnecessary patient transfers. Quinley *et al.* [26] evaluate use of smartphone photo-based inspection of the cervix for cervical cancer screening. Jaiganesh *et al.* [27] demonstrate that its is possible to diagnose Hamman's crunch from the sounds of one's breathing as recoded and analyzed on mobile phone.

Mitchell *et al.* [28] evaluated that the accuracy of diagnosis made with a smartphone based teleradiology system (employing the Non-Contrast-CT (NCCT) brain scans and Computerized Tomographic Angiography (CTA) head scans) for acute stroke are comparable to the ones conducted with a workstation. The smartphone teleradiology system appears promising and may have the potential to allow urgent management decisions in the case of an acute stroke. Similarly, Takao *et al.* [29] develop and successfully evaluate iStroke – a system using a smartphone for diagnostic image display and treatment of stroke, *i.e.*, exchanging clinical data, Computed Tomography (CT), Magnetic Resonance (MR), angio-graphic, intraoperative images, and expert opinion in real time.

Chun *et al.* [30] develop a more complex system, *i.e.*, a wrist-worn integrated health monitoring device (WIHMD) which performs the measurements of non-invasive blood pressure (NIBP), pulse oximetry (SpO2), electrocardiogram (ECG), respiration rate, heart rate, and body surface temperature and facilitates the detection of falls to determine the onset of emergency situation. The WIHMD also analyzes the acquired bio-signals and transmits the resultant data to a healthcare service centre through a smartphone. The authors have evaluated the system in real patient settings and they indicate that the system can be used as an effective tool for personalized diagnosis. Similarly, Oresko *et al.* [31] developed a wearable smartphone-based platform for real-time cardiovascular disease detection via electrocardiogram processing, where the smartphone is capable of performing real-time ECG acquisition and display, feature extraction, and beat classification. Jin *et al.* [32] similarly demonstrates ways of predicting cardiovascular disease from real-time electrocardiographic monitoring, using adaptive machine learning techniques deployed on a smartphone.

Wang *et al.* [33] report a simple and inexpensive microchip ELISA-based detection module that employs a portable detection system, *i.e.*, a cell phone-coupled device to quantify an ovarian cancer biomarker in urine. Integration of a mobile application immediately processing the microchip results eliminates the need for a bulky, expensive spectrophotometer. Similarly, Zhu *et al.* [34] successfully demonstrate the integration of imaging cytometry and fluorescent microscopy on a cell phone using a compact, lightweight, and cost-effective optofluidic attachment that could be useful for rapid and sensitive imaging of bodily fluids for conducting various cell counts (*e.g.*, toward monitoring of HIV+ patients) or rare cell analysis, as well as for screening of water quality in remote and resource-poor settings. In their research Zhu *et al.* [35] also demonstrate feasibility of a wide-field fluorescent microscopy on a smartphone in a case of labelling white-blood cells, as well as water-borne pathogenic protozoan parasites such as Giardia Lamblia cysts.

Stedtfeld *et al.* [36] present effectiveness, reproducibility and sensitivity level of Gene-Z - a device for point of care genetic testing that is using a smartphone processing capabilities to derive its results. Shen *et al.* [37] develop a standalone, wireless sensor integrating silicon nanowire field effect transistor, microfluidics and air sampling techniques for real-time monitoring biological aerosols. The authors demonstrate that, given an adequate smartphone-based application, the sensor is able to discriminate between H1N1 viruses and house dust allergens. Furthermore, Madan *et al.* [38] develop a smartphone application that predicts epidemiological trends (*e.g.*, running nose, sore throat and fever and even stress level and sadness) from mobile phone usage, mobility and time of a day and length of social interactions patterns. The authors conduct on a large user study, where some correlations have been identified, which however must be further tested for causality in more controlled settings.

### 3.2 Interventions/Treatments

Besides being a possible diagnosis tool, smartphone with its applications can support an ongoing chronic treatment or enable new acute treatment procedures, as we demonstrate in the following sub-sections.

#### 3.2.1 Chronic Case

The chronic treatment procedures include repeated interventions designed to improve particular, pre-defined patient's health outcomes. The smartphone is a platform suitable to support these procedures, as it is a personal device and it assists its user throughout different daily life activities and environments persistently. For example, Patrick *et al.* [39] design and successfully evaluate in a randomized controlled trial, a smartphone-based message-based intervention for a weight loss. The results show that messages on a phone might prove to be a productive channel of communication to promote behaviours that support weight loss in overweight adults. Similarly, Lee *et al.* [40] evaluate SmartDiet - a mobile phone-based diet game for weight control, and show that fat mass, weight and body mass index decrease significantly in the intervention group which acquires up-to-date nutrition information and manages its diet process via a smartphone application. Moreover, Free *et al.* [41] develop and validate the message-based application called txt2stop embracing the smoking cessation programme. The application demonstrates significant smoking cessation rates at 6 months and the authors recommend it to be considered for inclusion in services helping people to quit smoking. Moreover, Vogel *et al.* [42] demonstrate that anxiety and obsessive-compulsive disorder treatment show effectiveness when adaptive videoconference- and cell phone-based cognitive-behavioural therapy is employed. Amongst other smartphone-based repeated interventions we identify work of Kharbanda *et al.* [43] where the app called Text4Health enable an assessment of a parental readiness for text message immunization reminders, as well as work of Hardy *et al.* [44] conducting a randomized controlled trial of a personalized phone-based reminder system ARemind that prove to enhance adherence to antiretroviral therapy. Generalizing the approach, Clough and Casey [45] propose leveraging features of smartphones as a tool enabling better adherence to therapies. On the other hand, Lawton *et al.* [46] aim to empower the patients and decrease the risk of adverse drug events via a development of a personalized informational portal.

Given particular chronic condition, Quinn *et al.* [47] develop smartphone-based diabetes intervention study including personalized treatment via a communication between patients and practitioners. The results show that such a system may improve patient outcomes and be satisfactory to patients and physicians. The same authors also conduct cluster-randomized trial of a smartphone personalized behavioural intervention for blood glucose control [48]. The results show that the combination of behavioural mobile coaching with blood glucose data, lifestyle behaviours, and patient self-management data individually analyzed and presented with evidence-based guidelines to providers, substantially improve patient outcomes (*i.e.*, reduce glycated hemoglobin levels) over 1 year. Moreover, Morikawa *et al.* [49] study effects of salt reduction intervention program using an electronic salt sensor and smartphone on blood pressure among hypertensive workers. After 4 weeks, a greater decrease of blood pressure is observed in the in-tervention group, with significant reductions to daily salt excretion and blood pressure.

Kristjansdottir *et al.* [50] design an online situational feedback via a smartphone, to support self-management of chronic widespread pain. The intervention includes daily online entries (pain, daily life activities) and individualized message-based feedback from the therapist, grounded in a mindfulness-based cognitive behavioural approach. The intervention is rated as supportive, meaningful and user-friendly by the majority of the participants and overall is evaluated as successful.

Singh *et al.* [51] develop and evaluate mobile-phone based tele-dermatology system to support self-management of patients suffering from psoriasis – a skin condition where continuous clinical monitoring with periodic assessment of the state of the disease is essential for long-term therapy optimization. Photos and text describing observations are efficiently shared from a smartphone of a patient with a remote practitioner, which then can decide to admit the patient to the healthcare centre for further examinations.

Furthermore, Depp *et al.* [52] research mobile interventions for severe mental illness (*e.g.*, schizophrenia) by incorporating personal diary, messages and between-session mobile phone contacts with therapists. The results indicate that the mobile devices seem feasible and acceptable in augmenting psychosocial interventions for severe mental illness; however, future research is needed establishing the efficacy, cost effectiveness, and ethical and safety protocols for the proposed interventions. On the other hand, Puiatti *et al.* [53] develop a smartphone-centred wearable sensors network called MONARCA for monitoring patients with bipolar disorder. The system is meant to recognize early warning signs and predict maniac or depressive episodes.

On the spectrum of disabilities, Svoboda and Richards [54] aim at compensating for anterograde amnesia with use of theory-driven training program of technology use for individuals with moderate-to-severe memory impairment, where a smartphone is an object of intervention, as well as an assessment tool, while Kramer *et al.* [55] aim to empower blind users via developing face recognition tool based on a pictures captured by a camera of a smartphone.

### 3.2.2 Acute Case

As the smartphone is prevalent throughout different situations of daily life, it is also most likely to be found in cases, where acute intervention/treatment is required for instance to safe human life, but no specialized device is available. For example, Focosi [56] enumerate smartphone utilities for infectious diseases specialists in case of emergency, which allow them to identify chemical and biological hazards on the basis of reported symptoms and signs, a thesaurus helping them to make a differential diagnosis or an electronic version of compendium in the domain, as well as real-time translators, and finally a Beastmaster app (by PocketKai [57]) that can make a smartphone produce sounds of frequencies useful to expel gnats, fleas, house gnats, mice, rats, martens, and cockroaches (7-20 KHz).

Wu *et al.* [58] provide evidence of statistical association between the use of mobile phones to alert ambulance services in life-threatening situations and improved outcomes for patients, for example mobile phone compared to landline reporting of emergencies result in significant reductions in the risk of death at the scene. It is mainly because the use of mobile phones has the advantage of immediacy of access in particular in situations such as road traffic incidents, outdoor accidents, and injuries as well as incidents occurring at rural locations. The ambulance services can be alerted and react within a „golden hour".

Acute treatment supported by smartphone apps may also be conducted in case of out-of-hospital onsets of life threatening situations. Sikka *et al.* [59] present opportunities of the use of mobile phones for acute wound care (sending photos of wound and getting immediate response for what treatment is required). Paal *et al.* [60] is developing mobile-based basic life support (BLS) with metronome (producing regular, metrical beats per minute), to be used by lay rescuers for example in onset of cardiac arrest. Similarly, Ringh *et al.* [61] successfully demonstrate a case of use mobile phone technology to identify and recruit trained citizens to perform cardiopulmonary resuscitation on out-of-hospital cardiac arrest victims prior to ambulance arrival, increasing their chances of survival. Also, Scholten *et al.* [62] employs mobile phones to recruit laypersons for early cardiopulmonary resuscitation and use of automated external defibrillators in out-of-hospital cardiac arrest. Additionally, Bolle *et al.* [63] found that visual contact and supervision through video calls improve layperson rescuers' confidence in stressful emergencies. Looking boarder, Magee *et al.* [64] investigates possibility of using citizens' smartphones to get necessary actions on healthcare side, as well as policy side along public health critical events.

On the other side of the spectrum, we find a convenience factor as dominating in unconventional solutions postulating a use of smartphone in acute care. Examples include Peters *et al.* [65] developing an smartphone app using accelerometer and camera and improving the surgeons' ability to correct the acetabular cup orientation in total hip arthroplasty, and Brusco [66] who successfully employs smartphone applications in perioperative practice, providing relevant patient information to the surgeon before, along and after the operation and improving his confidence.

### 3.3 Rehabilitation

Maddison *et al.* [67] develop HEART system being evaluated in a randomized controlled trial study protocol for cardiac rehabilitation (CR), aiming at improving health behaviours to slow or reverse the progression of cardio-vascular disease (CVD). The intervention consists of a theory-based, personalized, automated package of text and video message components via participants' smartphones and it prove to statistically significantly increase participants' exercise behaviour. Additionally, Worringham *et al.* [68] develop and prove feasibility of a smartphone, Electrocardiography (ECG) and Global Positioning System (GPS) based system to remotely monitor exercise behaviours in CR. The system provides a feasible and very flexible alternative form of supervised cardiac rehabilitation for those unable to access hospital-based programs. Piotrowicz *et al.* [69] evaluate home cardiac rehabilitation, where the patient is telemonitored with an ECG device, transmitting its data via mobile phone to a monitoring centre. Similarly, Marshall *et al.* [70] develops a smartphone application for improved self-management of pulmonary rehabilitation, including a remote participation in a therapeutics-supervised rehabilitative exercise programme.

## 3.4 Education and Training

Smartphone technologies facilitate also information and communication services, that can be used in education and training of future practitioners. As concrete examples, work of Phillippi and Wyatt [71] discusses use of smartphones in nursing education, for a quick access to educational materials and guidelines during clinical procedures, classes, or clinical conferences. Students can review instructional videos prior to performing skills and readily reach their clinical instructor via message system. Downloadable applications, subscriptions, and reference materials expand the smartphone functions even further.

Abboudi and Amin [72] and Makanjuola and Bultitude [73] identify 200+ surgical applications for the urology trainee, nurse or a patient, including contributions from a range of specialties such as plastic, breast, orthopaedic, opthalmological, cardiothoracic, colorectal and maxillo facial surgery. They indicate that smartphones are now capable of storing high quality ultrasonography and CT and, coupled with the high resolution cameras, mobile radiology can be used to gain rapid opinions from senior colleagues when trainees are off site, allowing early intervention and potentially better patient outcomes.

Franco and Tirrell [74, 75] identify 61 iPhone and 13 Android apps for orthopaedic surgeons. Yet, the authors admit that only few highly ranked apps specifically related to orthopaedic surgery are available, and the types of apps available (technique guides, reference, and industry/news) do not appear to be the categories most desired by residents and surgeons (textbook/reference, techniques/guides, Orthopaedics In-Training Exam/board review, and billing/coding). Amin [76] surveys smartphone applications for the plastic surgery trainee. The results indicate that there are currently very few plastic surgery applications available useful to the plastic surgeon or a junior plastic surgery trainee, but there exist remote consultations with access to digital images, allowing triaging and resulting in an improved level of care, as ultrasound scans and computed tomography scans can now be stored on the phone itself. Particular example of software enabling a trainee as well as a practitioner to access digital radiology images is OsiriX, which is evaluated by Choudri *et al.* [77] as a tool for communicating between specialists, and as a training tool for surgeons and in conjunction with augmented reality techniques by Volonte *et al.* [78]. In general, Dala-Ali *et al.* [79] suggest a rise in general use of the smartphone platform for surgeons, enabling them to download the medical news, trainings, podcasts, textbooks, drug databases, and more. Pope *et al.* [80] discuss smartphone use for the 'modern day otolaryngologist', including news, medical info, useful formulas and equations, compendium of drugs and doses, and apps for assess an individual's hearing, including 3D animations of common conditions and their treatment. Brunet *et al.* [81] emphasize the importance of use of ICT at large, and role of personal smartphones podcasting of lectures for students of medical school.

With increased information, there is also an increased need to appropriate mobile search applications. As Muller *et al.* [82] indicate the quality of the information is a very important. In their research, they explore aspects of medical information search, text retrieval using ontologies for semantic text analysis and visual retrieval, to include images and videos. The related research project, as presented by Hanbury *et al.* [83], aims at using semantic, textual and visual information retrieval targeted at three main groups: the general population, GPs and radiologists. In their recent paper, Depeursinge *et al.* [84] prove usability of an application enabling a mobile access to peer-reviewed medical information based on textual search and content-based visual image retrieval.

## 3.5 Personal Integrated Health Informatics Systems and Apps

As we present in the followings section, there exists an increasing body of developments in personal health informatics systems and applications, which particularly target a personal smartphone as a delivery platform. On one side, we witness Harvard Womens Health Watch [85] suggesting credible apps that can be used for personal health, and this list was specifically compiled with their female patients in mind. Therefore its suggests a) the example MyOBGYN and Medscape apps, that provide state-of-the-art medical information or the latest medical news; b) apps that act as period trackers; c) apps that help with the exercises that strengthen the pelvic floor muscle to prevent urine from leaking Kegel); d) apps that facilitate food and sports activities diaries and provide a social space for building supportive communities for weight management; e) apps that help with medication adherence; f) public health info apps that keep one informed on foodborne diseases and potential pandemics.

Moreover, there exist examples of personalized health informatics solutions, that are successfully adopted in wellness domain (sports and fitness) and that rely on sensor informatics technologies built-in or connected to a smartphone, like: a) AirStrip – heart rate (HR), pressure, body temperature, oxygen saturation in the blood sensors [86]; b) „basis" device - wristwatch optical sensor, accelerometer, galvanic skin response (GSR), skin/ambient temperature [87]; c) Health Buddy – smartphone connected blood pressure and thermometer [88]; d) Zio iRhythm ECG patch (holter monitoring) [89]; e) corventis – PiiX patch for ECG, HR, fluid status, oxygen, respiration, temperature, current position for activity [90]; f) minimized Paradigm Real-time Revel System from Medtronic combining an insulin pump with regular monitoring of glucose levels for diabetes management [91]; g) respiration belt by Kai Medical – respiratory rate, patterns, and activity [92]; h) Zeo monitoring electroencephalography (EEG) in sleep [93]; i) SleepMinder by BlancaMed using motion sensors to measure sleep quality, respiration, and any incidents of sleep apnea without contact with the body [94]; j) Lark measuring micro-motions for sleep patterns [95]; k) Proteus, microsensor-enabled medication to track medication compliance and customize therapy [96].

Particularly for fitness we consider the following examples of personalized sensors and corresponding smartphone apps: a) Nike+ - shoe GPS, and chest band for HR [97]; b) Wahoo - HR chestband [98]; c) Whinings - blood pressure and body scale estimating body fat and body lean mass (through known equations for the density of fat and fat free mass) [99]; d) Runmeter - external GPS tracking [100]; e) Philips direct life - external accelerometer [101]; f) Endomondo - external GPS [102]; g) Motorola MotoActv - external GPS [103]; h) FitBit - external accelerometer [104]; i) Garmin watch - external GPS and HR [105]; j) BodyMedia SenseWear - external thermometer, GSR and accelerometer [106]; k) Jawbone UP - external wrist band accelerometer [107].

On a side of research platforms for wellness and fitness we indicate a) own Activity Level Estimator – relying on smartphone built-in accelerometer [108, 109]; b) use of a smartphone camera for photoplethysmography from which Jonathan and Leahy [110] derive change in the heart rate; c) a portable ultrasound pulsed-wave (PW) Doppler flowmeter using a smartphone by Huang *et al.* [111]; or d) work of Wu *et al.* [112] aiming at recognizing mobility activities using a BlackBerry smartphone's built-in accelerometer, GPS, video camera, and timer to identify static activities, walking-related activities, *etc.*

Nutrition intake is an important component of lifestyle. Arab *et al.* [113] as well as Chen *et al.* [114] focus on smartphone-based nutrition app that enables a user to take photo images of foods they consumed, and they both show that automated imaging is a promising technology to facilitate dietary recall. Chen *et al.* additionally propose smartphone-based software that automatically analyzes the photos to recognize dishes and estimate calories. Moreover Chae *et al.* [115] evaluate smartphone app that, based on photos of food, automatically estimates food volumes through the use of shape templates.

On the other side of the personalized apps spectrum is an integrated system, which although in a fixed location, may be available for many persons. It is called Health Capsule (by Chengdu Internet of Things Technology Institute (CITTI)) and it is an (kiosk-like) integrated system operational in, *e.g.*, shopping centre enabling a low-cost ad-hoc checkups and on-demand online diagnosis and medication prescription [116].

# 4 Discussion

The overview presented in this paper has several limitations, especially with respect to the number of articles considered and limited areas of health services. The article therefore exhibits limitations with respect to the extent, to which the findings can be generalized beyond the cases studied. Nevertheless, the research presented in this paper reveals several shortcomings of the current smartphone-driven developments in health services provision. A prevailing aspect indicated by several authors is lack of evidence-based applications, and lack of involvement of healthcare providers in developing these applications. As concrete example, results from study of Gan and Allman-Farinelli [117], auditing 403 smartphone applications for the management of obesity indicate, that only a very small amount of applications (exactly 8) is evidence-based and can be recommended as useful adjunct treatment to health professionals' advice, assisting their patients' weight loss efforts. Out of eight applications, there are five calorie and physical activity counters types of applications and three Body-Mass-Index (BMI) or weight trackers types of applications. Without a practitioner support, the authors indicate that there is a large responsibility of application choice on its user.

Similarly, Rosser and Eccleston [118] audit 11 smartphone applications for pain self-management, and conclude that there is a low level of stated health-care professionals' involvement in app development and content. The critic is that the apps appear to be able to promise pain relief without any concern for the effectiveness of the product, or for possible adverse effects of service use. Given a population of pain sufferers is being often desperate for a solution to their distressing and debilitating pain conditions, the authors point out that there is considerable risk of individuals being misled when using these or similar applications accessible on their smartphones.

With respect to the disease self-management, there are also positive developments, driven by liability concerns of app developers and policy-makers. For instance, in July 2010, the US-based Food and Drug Administration (FDA) ruled that the WellDoc Diabetes Manager System [119], designed for self-management in diabetics, should be marketed as a medical device because it not only logs glucose levels but also gives medical advice based on the results. The certified app is available since 2011; apps for asthma, cardiovascular disease, and cancer are scheduled to follow.

On a side of practitioners-targeted applications, shortcomings are identified as well. For example Stanzel [120], when auditing 70 smartphone applications aimed at ophthalmology practitioners, conclude that many of them lack credibility and have low usability, as they are unreliable (and crash) when operational on a smartphone. Additionally, Visvanathan [121], reviewing 94 microbiology-themed apps, 'reference' (microbiology textbooks, laboratory/diagnostic test interpretations, guidelines), 'education' (microbiology questions/ûashcards for examinations, educational talks), most popular 'antibiotic guidance' (pharmacology advice, dose calculators) for diagnosis and patient management, raise concerns specifically regarding accuracy and reliability of applications' content. Some applications' content is biased towards particular industry-providers, which is undesirable.

Furthermore, Hamilton and Brady [122] conduct a survey of 79 dermatology-themed smartphone apps in which they specifically examine the authorship of the app, in order to gauge the prevalence of medical professional involvement in app development and content. They reason that it would seem essential that expert medical personnel is involved in the creation of most medical apps, however they demonstrate differently, and they additionally emphasize that that certain apps provide many potential diagnostic risks for the untrained user. The authors suggest an increased regulation to improve accountability of apps content.

Summarizing our overview's results, we conclude that current smartphone applications lack to the large extend a solid evidence-based approach and assurance of the credibility of the content, *i.e.*, they lack in their design the conscientious use of current best evidence in making suggestions about the care of their user or the delivery of health services to their user. Visvanathan [123] further suggests that phones are an efficient, common, and popular means of communication, however all users in the clinical environment urgently require clear evidence-based guidelines to avoid potential pitfalls.

## 5 Conclusions and Outlook

Personal sensors, signal and imaging informatics exist, but still have a long way to go to become an everyday, personalized healthcare-provisioning tool in the medical field and in a clinical practice. The critical technological advances and user acceptance witnessed in the self-care can pave the way. Key challenges for their widespread adoption involve, as identified in the above-mentioned literature, lack of user acceptance striving from variable credibility and reliability of applications and solutions as they a) lack evidence-based approach; b) have low levels of medical professional involvement in their design and content; c) are provided in an unreliable way, influencing negatively its usability; and, in some cases, d) being industry-driven, hence exposing bias in information provided, including bias towards particular types of treatment or intervention procedures.

According to the literature, particularly the awareness of healthcare practitioners, and their willingness in being involved in developing new generation of applications and tools, may be a driving force for adoption. There is a hope in that domain. Namely, as Franko and Tirrell [75] indicate, the smartphone app use among medical providers is high, 85% of their respondents used a smartphone, of which the iPhone was the most popular (56%). Over half of the respondents reported using apps in their clinical practice; the most commonly used app types were drug guides (79%), medical calculators (18%), coding and billing apps (4%) and pregnancy wheels (4%). The most frequently requested app types were textbook/reference materials (average response: 55%), classification/treatment algorithms (46%) and general medical knowledge (43%). The clinical use of smartphones and apps will likely continue to increase, and the authors indicate an absence of high-quality and popular apps despite a strong desire among physicians and trainees. Similarly, Choi *et al.* [124] documented a use of the smartphone for doctors in an empirical study from Samsung medical centre, where they developed a Dr SMART app that gives doctors mobile access to patient information. The most commonly accessed content was inpatient information; this constituted 78.6% of all accesses, within this 50% was to accesses lab results.

Menon [125] is a medical practitioner in emergency medicine that confessed openly „I Can't Live Without My Smartphone" and gives a practical usage of this tool in a clinical care. His question is provoking „ if you do not have a smartphone, then ask yourself this: Can I practice medicine with-out it?" and we think that if every practitioner and patient would attempt to answer it, a critical mass will pave a new way to advances towards personal health informatics. On the other hand, Mertz [126] claims „there is an app for everything". Engineers, computer programmers, medical professionals, and other researchers are abundantly creating apps and add-on devices, or peripherals, that turn a smart phone into a microscope, an ultrasound machine, or a heart-rate monitor, just to name a few. But the key was, is and always will be the quality and credibility of information, influencing the quality of health services delivery. The next challenge lies there.

**Acknowledgements**

The author is partially supported by the EU AAL research projects TraiNutri (09-2-129), WayFiS (10-3-014) and MyGuardian (11-4). The author thanks the reviewers for their constructive comments for improving the paper structure and content.

**Correspondence to:**
Katarzyna Wac
University of Geneva
Institute of Services Science
Quality of Life Group
Geneva, Switzerland
E-mail: katarzyna.wac@unige.ch